# TIME INTERVAL AND LATTICE VIBRATION IN RAMAN EFFECT, PHOTOELECTRIC EFFECT AND PLANE MIRROR REFLECTION

**M. Kumar and S. Sahoo[1]**

Department of Physics, National Institute of Technology,

Durgapur – 713209, West Bengal, India.

[1]E-mail: sukadevsahoo@yahoo.com

## Abstract

Time interval between the incident and scattered photon in Raman effect and absorption of photon and emission of electron in photoelectric effect has not been determined till now. This is because there is no such high level instrument discovered till now to detect time interval to such a small level. But this can be calculated theoretically by applying a basic principle of physics like impulse is equal to the change in momentum. Considering the collision between electron and photon as perfect inelastic collision in photoelectric effect, elastic and inelastic collision in Raman effect and elastic collision in plane mirror reflection and the interaction between electron and photon as strong gravitational interaction we calculate the required time interval. During these phenomena there is lattice vibration which can be quantized as phonon particles.

**Keywords:** Photon absorption and scattering; Phonon; Strong gravity

**PACS:** 25.20.Dc; 63.20.-e

## 1. Introduction

In photoelectric effect [1–4], electrons are emitted from matter as a consequence of their absorption of energy from electromagnetic radiation, such as ultraviolet radiation or x-rays. Many attempts have been made to measure the time taken by electrons to absorb enough energy from the light to escape from the matter. But no detectable time lag has been found so far [5,6]. According to the photon interpretation of the photoelectric effect, there will be no appreciable time lag between light striking the surface of the matter and the start of photoemission. The electrons absorb the energy from the light beam nearly instantaneously. From experiments Atkinson [7] showed that the excitation of an atom takes place in less than $10^{-10}$ s. Brandl [6] theoretically calculated the absorption time is roughly $1/3 \times 10^{-22}$ s. But such a short time interval could not be measured experimentally so far. Therefore it is of special interest to study such phenomenon both theoretically as well as experimentally.

The Raman effect [8–12] has a tremendous impact on the natural sciences. In Raman effect, photons are scattered from an atom or a molecule. Due to the interaction of light with different elementary excitations in solids, Raman spectroscopy is used to study different solid state properties like lattice vibration (phonons), charge carrier dynamics and electronic structure, spin dynamics and magnetic order (magnons) [13–15]. Recently [16], Raman effect is discussed in femto- and attosecond time scales. But the time interval between incident and scattered photon in Raman effect is not measured so far. Similarly when light is incident on a plane mirror, it is reflected back obeying the rules of reflection [3]. The time interval between interaction of photon and electron in plane mirror reflection is also not measured so far.

In this paper, we calculate time interval between the incident and scattered photon in Raman effect, time interval between the absorption of photon and emission of electron in photoelectric effect and the time interval between the interaction of photon and electron in plane mirror reflection theoretically by considering the interaction between electron and photon as strong gravitational interaction [17,18]. Generally strong gravity acts at the level of elementary particles. Unlike the universal gravitational constant G, the strong gravitational constant $\Gamma$ depends on the type of objects. Tennakone [17] has treated the electron and the proton as black holes in the strong gravitational field and calculated the strong gravitational constant as: $\Gamma = 3.9 \times 10^{28}$ m$^3$kg$^{-1}$s$^{-2}$. In [18], the author has found $\Gamma = 2.77 \times 10^{32}$ m$^3$kg$^{-1}$s$^{-2}$ by studying strong gravitation which plays an important role in the stability of elementary particles.

This paper is organized as follows: by considering the interaction between electron and photon as strong gravitational interaction, we calculate the time interval between the incident and scattered photon in Raman effect in Section 2. In Section 3, we calculate the time interval between absorption of photon and emission of electron in photoelectric effect. In Section 4, we calculate the time interval between interaction of photon and electron in plane mirror reflection. During these phenomena there is lattice vibration which can be quantized in terms of phonon particles. Then we present our conclusions in Section 5.

## 2. Time interval between incident and scattered photon and lattice vibration in Raman effect

Electrostatic energy of electron is finite because radius of electron is finite [17]. The extension of charge distribution inside the electron which interacts with each other are not able to hold together. In order to prevent the explosion of the electron, non-electromagnetic forces are required. To overcome the electrostatic repulsion among the distributed charge inside the electron, strong gravitational interaction is required. Considering the interaction between electron and photon as strong gravitational interaction, we can calculate the time interval between the incident and scattered photon in Raman effect [19]. In Raman effect, the photon is scattered by an electron of the atom. As we know impulse is equal to the change in momentum of the photon which can be given by the equation

$$\frac{\Gamma m_e \frac{h\nu_1}{c^2} \Delta t}{R_e^2} = \frac{h\nu_2}{c} + \frac{h\nu_1}{c} \tag{1}$$

In equation (1), $\Gamma$ represents the strong gravitational constant, $m_e$ represents the mass of the electron, $\nu_1, \nu_2$ are the frequencies of photons incident and scattered respectively, $\Delta t$ is the time interval between the incident and scattered photon, $R_e$ is the radius of the electron, $h$ is Planck's constant and c is the speed of light. In the LHS of equation (1), $\Gamma m_e \dfrac{h\nu_1}{c^2}/R_e^2$ represents the strong gravitational force between electron and photon and $\dfrac{\Gamma m_e \dfrac{h\nu_1}{c^2} \Delta t}{R_e^2}$ represents the impulse.

The RHS of equation (1) represents the change in momentum of photon $(h\nu_2/c - (-h\nu_1/c))$. Here we consider the maximum scattering of incident photon i.e. by $180^0$. Equation (1) can be written as:

$$\frac{\Gamma m_e \Delta t}{\lambda_1 c R_e^2} = \frac{1}{\lambda_2} + \frac{1}{\lambda_1} \tag{2}$$

where $\lambda_1$ and $\lambda_2$ are wavelengths of the incident and scattered photon respectively. From equation (2) we get

$$\Delta t = \frac{c R_e^2}{\Gamma m_e}(1 + \frac{\lambda_1}{\lambda_2}) \tag{3}$$

Using $\Gamma = 2.77 \times 10^{32} \, m^3 kg^{-1} s^{-2}$ [18], for X-rays $\lambda_1 = 1 A^0, \lambda_2 = 100 A^0$ [20], $R_e = 2.82 \times 10^{-15} m$ and $m_e = 9.11 \times 10^{-31} kg$ [21] we get $\Delta t = 0.9464 \times 10^{-23} s$. Again for $\lambda_1 = 100 A^0$ and $\lambda_2 = 0.1 A^0$ we get $\Delta t = 0.9463454 \times 10^{-20} s$. i.e. in Raman effect, when the photon gives energy to the electron, time interval of interaction is less but when it takes energy, time interval of interaction between electron and photon is more. For elastic collision $\lambda_1 = \lambda_2$ and we get $\Delta t = 1.8908 \times 10^{-23} s$.

During this phenomenon there is lattice vibration which can be quantized as phonon particles [22]. Applying conservation of momentum for the three particles i.e. incident photon, scattered photon and the phonon, equation can be written as

$$-\frac{h}{\lambda_1} = \frac{h}{\lambda_2} \pm \frac{h}{\lambda_P}, \tag{4}$$

where $\lambda_p$ is the wavelength of the phonon particle. Equation (4) can be simplified into

$$-\frac{1}{\lambda_1} = \frac{1}{\lambda_2} \pm \frac{1}{\lambda_P} \tag{5}$$

In equations (4) and (5) ± sign indicates the creation and absorption of phonon particle in the lattice vibration respectively.

## 3. Time interval between absorption of photon and emission of electron and lattice vibration in photoelectric effect

In the same way considering strong gravitational interaction between electron and photon in photoelectric effect, we can calculate the time interval between absorption of photon and emission of electron in photoelectric effect. We know that impulse is equal to the change in momentum, which can be written as:

$$\frac{\Gamma m_e \frac{h\nu}{c^2} \Delta t}{R_e^2} = m_e v + \frac{h\nu}{c} \quad (6)$$

In equation (6), v is the velocity of emitted electron, $\nu$ is the frequency of photon absorbed. In the LHS of equation (6), $\Gamma m_e \frac{h\nu}{c^2} / R_e^2$ represents the strong gravitational force between electron and photon and $\frac{\Gamma m_e \frac{h\nu}{c^2} \Delta t}{R_e^2}$ represents the impulse. The RHS of equation (6) represents the change in momentum $(m_e v - (-h\nu/c))$. For $\lambda = 4500 A^0$ and $v = 4 \times 10^5 m/s$ [21], we get $\Delta t = 2.349838 \times 10^{-21} s$. From equation (6), it is clear that $\Delta t$ depends upon the velocity of emitted electron and frequency of absorbed photon. We change the frequency from infrared to gamma rays i.e. $10^{14}$ to $10^{22}$ Hz and velocity from 0 to ~$10^6$ m/s we get $\Delta t$ varies in between $0.9454 \times 10^{-23}$ s to $1.6463 \times 10^{-20}$ s. In photoelectric effect, time interval of interaction between electron and photon decreases with the increase in frequency or energy of the photon.

During this phenomenon there is lattice vibration which can be quantized as phonon particles [22]. Applying conservation of momentum for the three particles i.e. absorbed photon, emitted electron and the phonon, equation can be given below

$$-\frac{h}{\lambda} = m_e v \pm \frac{h}{\lambda_P} \quad (7)$$

Here, ± sign indicates the creation and absorption of phonon particle in the lattice vibration respectively.

## 4. Time interval between interaction of photon and electron and lattice vibration in plane mirror reflection

Light incident on plane mirror is reflected back following the rules of reflection i.e. angle of incidence is equal to the angle of reflection and the incident light, reflected light and the perpendicular lie on the same plane. If the incident light is along the perpendicular, it is reflected back along the perpendicular. Considering the interaction between electron and photon as strong gravitational interaction on the surface of the plane mirror, we can calculate the time interval between interaction of photon and electron in plane mirror reflection. As we know impulse is equal to the change in momentum of the photon which can be given by the equation

$$\frac{\Gamma m_e \frac{h\nu}{c^2} \Delta t}{R_e^2} = \frac{2h\nu}{c}. \tag{8}$$

In the LHS of equation (8), $\Gamma m_e \frac{h\nu}{c^2} / R_e^2$ represents the strong gravitational force between electron and photon and $\frac{\Gamma m_e \frac{h\nu}{c^2} \Delta t}{R_e^2}$ represents the impulse. The RHS of equation (8) represents the change in momentum $(h\nu/c - (-h\nu/c))$. Solving equation (8), we get the value of $\Delta t$ as given below

$$\Delta t = \frac{2cR_e^2}{\Gamma m_e}. \tag{9}$$

From equation (9), it is clear that time interval in interaction between electron and photon in plane mirror reflection is a constant quantity and is independent of the frequency of the incident light. Putting the values of $c$, $R_e$, $\Gamma$, $m_e$, in equation (9), we get $\Delta t = 1.8908 \times 10^{-23} s$.

During this phenomenon there is lattice vibration which can be quantized as phonon particles. Applying conservation of momentum for the three particles i.e. incident photon, reflected photon and the phonon we can write

$$\frac{2h}{\lambda} = \pm \frac{h}{\lambda_P} \tag{10}$$

Equation (10) can be solved to give the value of $\lambda_P$ as,

$$\lambda_P = \pm \frac{\lambda}{2} \tag{11}$$

In equations (10) and (11) $\pm$ sign indicates the creation and absorption of phonon particle in the lattice vibration respectively.

## 5. Conclusions

Finally, we have come to the conclusion that the time interval between incident and scattered photon in Raman effect and the time interval between absorbed photon and emitted electron in photoelectric effect varies from $\sim 10^{-23} s$ to $\sim 10^{-20} s$ which is calculated by considering the interaction between electron and photon as strong gravitational interaction. In case of plane mirror reflection, interaction between electron and photon on the surface of the mirror occurs for the time interval of the order of $10^{-23} s$. The result of the above phenomena is lattice vibration which can be quantized as phonon particles. Brandl [6] theoretically calculated the absorption time for photoelectric effect is roughly $1/3 \times 10^{-22}$ s, which is within the range of our results. But such a short time interval could not be measured experimentally so far. Therefore it is of special interest to study these phenomena both theoretically as well as experimentally.

## Acknowledgments

We would like to thank P. K. Mohanty, TIFR, Mumbai; P. K. Swain, IIT, Kharagpur and J. K. Mohapatra, IIT, Mumbai for their help in the preparation of the manuscript. We thank the referee for suggesting valuable improvements of our manuscript.

## References

1. A. Beiser, S. Mahajan and S. R. Choudhury, *Concepts of Modern Physics*, Sixth Edition, Special Indian Edition, Tata McGraw-Hill, Delhi (2009).
2. R. A. Serway, C. J. Moses and C. A. Moyer, *Modern Physics*, Third Edition, Brooks/Cole, Belmond (2005).
3. F. W. Sears, M. W. Zemansky and H. D. Young, *University Physics*, Sixth Edition, Narosa Publishing House, New Delhi (1998).
4. H. Heinrich, *Annalen der Physik*, **267**(8), 983 (1887).
5. E. O. Lawrence and J. W. Beams, *Phys. Rev.* **32**, 478 (1928).
6. M. Brandl, Project Physnet-MISN-0-213 – The Photoelectric effect, 11/08/2001.
7. d'E. Atkinson, *Roy. Soc. Proc.* **A116**, 81 (1927).
8. C. V. Raman and K. S. Krishnan, *Nature*, **121**, 507 (1928).
9. C. V. Raman and K. S. Krishnan, *Nature*, **121**, 711 (1928).
10. C. V. Raman and K. S. Krishnan, *Nature*, **122**, 12 (1928).
11. C. V. Raman, *Nature*, **123**, 50 (1929).
12. G. Landsberg and L. Mandelstam, *Naturwissenschaften*, **16**, 557 (1928).
13. M. Opel and F. Venturi, *European Pharmaceutical Review*, **7**(3), 76 (2002).
14. R. Loudon, *Proc. Phys. Soc.,* **82**, 393 (1963).
15. R. Loudon, *Adv. Phys.* **50**(7), 813 (2001).
16. A. M. Zheltikov, *Phys.-Usp.,* **54**(1), 29 (2011).
17. K. Tennakone, *Phys. Rev. D*, **10**, 1722 (1974).
18. J. J. Perng, *Nuovo Cimento, Lettere*, Series 2, **23**(15), 552 (1978).
19. S. L. Gupta, V. Kumar and R. C. Sharma, *Elements of Spectroscopy*, Pragati Prakashan, Meerut (1996).
20. Frank S Crawford Jr., *Waves*, Tata McGraw-Hill, Indian Edition, New Delhi (2008).
21. S. N. Ghoshal, *Atomic Physics*, S. Chand & Company Ltd, New Delhi (2003).
22. C. Kittel, *Introduction to Solid State Physics*, 7th Edition, Wiley-India, New Delhi (2008).